\documentclass[preprint, a]{iucr}              

                   \def\href#1{\relax}\let\foo\caption
\ifPDF
  \RequirePackage{hyperref}
  \PassOptionsToPackage{pdftex,bookmarksopen,bookmarksnumbered}{hyperref}
\fi
\let\caption\foo

\usepackage{amsthm}
\usepackage{mathptmx}
\usepackage{amssymb}
\usepackage{amsfonts}
\usepackage{tikz}
\usepackage{mathptmx}
\usepackage{amsmath,amscd}
\usepackage{pgfplots}
\usepgflibrary{shapes.geometric} 
\usepgflibrary[shapes.geometric] 
\usetikzlibrary{shapes.geometric} 
\usetikzlibrary{arrows,shapes,backgrounds}

\usepackage{tikz}
\usepackage{pifont}

\usepackage{multicol}        

 \theoremstyle{definition}
 
 \theoremstyle{remark}

 \numberwithin{equation}{section}

     \paperprodcode{a000000}      
     \paperref{xx9999}            
     \papertype{FA}               

     \paperlang{english}          
     \journalcode{A}             
     \journalyr{2013}
     \journalreceived{\relax}
     \journalaccepted{\relax}
     \journalonline{\relax}

\begin{document}                  



\title{Viruses and Fullerenes \-- Symmetry as a Common Thread?}
\shorttitle{Carbon onions from affine extensions}


     \cauthor[a]{Pierre-Philippe}{Dechant}
{ppd22@cantab.net}{}

    \author[b]{Jess}{Wardman}

    \author[b]{Tom}{Keef}

    \author[b]{Reidun}{Twarock}

\aff[a]{Mathematics Department, Durham University, \city{Durham, DH1 3LE}, \country{United Kingdom}}    
   \aff[b]{Mathematics Department, University of York, \city{Heslington, York, YO10 5GG}, \country{United Kingdom}}


\shortauthor{Wardman, Dechant, Keef, Twarock}







\maketitle                        

\begin{synopsis}
Carbon onions from affine extensions in analogy with virus work.
\end{synopsis}

\begin{abstract}
We apply here the  principle of affine symmetry  to the nested fullerene cages (carbon onions) that arise in the context of carbon chemistry. Previous work on affine extensions of the icosahedral group has revealed a new organisational principle in virus structure and assembly. We adapt this group theoretic framework here to the physical requirements dictated by carbon chemistry, and show that we can derive mathematical models for carbon onions within this affine symmetry approach. 
This suggests the applicability of affine symmetry in a wider context in Nature, as well as offering a novel perspective on the geometric principles underpinning carbon chemistry.
\end{abstract}


%
	
\section{Introduction} \label{ACA_Intro}

Symmetry -- and in particular rotation and reflection symmetry --  features very prominently in the natural world, for instance in crystals and viruses.
Often these symmetric structures correspond to minimum energy configurations, or are driven by
other principles such as that of genetic economy in virology \cite{Crick:1956}. 
Whilst crystalline arrangements and their crystallographic reflection symmetries are undoubtedly very important, the largest reflection groups in two and three dimensions are actually non-crystallographic: the symmetries of the regular polygons in two dimensions $I_2(n)$, and the symmetry group of the icosahedron in three dimensions $H_3=I_h$. For example, in many cases the proteins in viral capsids are organised according to (rotational) icosahedral symmetry $I$. Thus symmetry is an important principle for virus structure, assembly and dynamics. 

Icosahedral arrangements of carbon atoms have also been observed since the 1980s, collectively known as fullerenes \cite{Ugarte:1995,Ugarte:1992,Hawkins:1991}. 
The most prominent icosahedral fullerene is the buckyball $C_{60}$ \cite{Kroto:1985}, which in mathematical nomenclature is called a truncated icosahedron and has the shape of a football. Larger such fullerene configurations also exist, and  of particular interest here are nested arrangements of fullerene cages, called carbon onions  \cite{Iijima:1980}. All these cages share the  property that carbon atoms each have three bonds to other carbon atoms of roughly the same length and angle, i.e. fullerene cages are three-connected.

A symmetry point group describes a structure at a given radial distance from the origin, such as a single fullerene cage. Via an affine extension, the symmetry point group is augmented so that it can describe structures at different radial levels collectively \cite{DechantTwarockBoehm2011H3aff,DechantTwarockBoehm2011E8A4,Dechant2012AGACSE}. It therefore lends itself to the modelling of carbon onions \cite{Twarock:2002a,Twarock:2002b}. Recently we introduced new affine extensions for the icosahedral group \cite{Twarock:2006b,Keef:2009} and demonstrated that these can be used to model virus structure at different radial levels \cite{Keef:2013,Wardman:2012}. In particular, this work revealed a previously unappreciated molecular scaling principle in virology, relating the structure of the viral capsid of Pariacoto virus to that of its packaged genome. This suggests that the overall organisation of such viruses follows  an affine version of the icosahedral group, and implications of this discovery for virus dynamics and assembly have been discussed based on this new principle \cite{Indelicato:2011}.


We will investigate here whether these new mathematical structures can also model carbon onions. In particular, we examine the possibility that the different shells of a carbon onion can collectively be modelled  via such an affine symmetry. We extend and adapt our earlier work on affine symmetry in the context of viruses to the physical situation dictated by the constraints of carbon chemistry.
That is, we investigate affine extensions of icosahedral symmetry that are compatible with three-connectedness of the carbon atoms in fullerenes.

This paper is organised as follows. Section \ref{ACA_MathBack} introduces and illustrates the principle of affine symmetry, how it is applied to viruses, and how we adapt it here to the context of carbon chemistry. 
Section \ref{ACA_CarbonOnions} demonstrates how the experimentally observed carbon onions follow straightforwardly from our affine symmetry framework. We conclude in Section \ref{ACA_Concl}.

\section{The Symmetry Paradigm} \label{ACA_MathBack}

\begin{figure}
\begin{center}
\includegraphics[height=5cm]{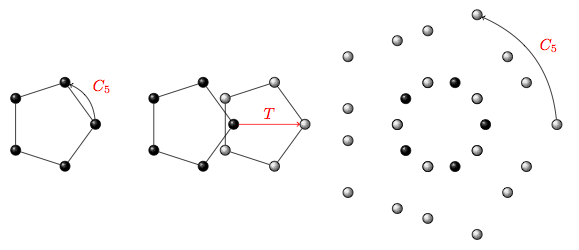}
\end{center}
\caption{A planar example of affine symmetry: the action of an affine extended symmetry group on a pentagon. The translation operator $T$, in combination with the rotational symmetry group $C_5$, generates multiple copies of the pentagon, with vertices corresponding to words in the generators of the extended group. Due to the action of $C_5$ the vertices form a $C_5$-symmetric point array. Coinciding points, i.e. points located on more than one pentagon,  correspond to distinguished translations and non-trivial group structures, which can be used to model natural phenomena.}
\label{figpent}
\end{figure}

The principle of affinisation, i.e. the extension of a finite symmetry group by the addition of a non-compact generator, is commonly used in the context of crystallographic groups to generate space groups. It has been introduced in a non-crystallographic setting for the first time in \cite{Twarock:2002a}. In particular, in this reference the reflection groups $H_3$ and $H_4$, which are the only reflection groups containing icosahedral symmetry as a subgroup, have been extended by an affine reflection, and it has been shown that in combination with generators of the finite groups the affine reflections act as translations. Subsequently, affinisations of icosahedral symmetry via translation operators have been classified \cite{Keef:2009,Keef:2013,DechantTwarockBoehm2011H3aff,DechantTwarockBoehm2011E8A4,Wardman:2012}. These contain the affinisations derived previously, but in addition provide a much wider spectrum of extended group structures. We illustrate the construction principle geometrically here for the  two-dimensional example of the rotational symmetry group $C_5$ of a regular pentagon, see Fig.~\ref{figpent}. 

Addition of the  translation operator $T$ (here taken to be of the same length as the golden ratio $\tau$ times the radius of the circle into which the pentagon is inscribed) makes the group non-compact by creating a displaced version of the original pentagon. The action of the symmetry group $C_5$ of the pentagon generates additional copies in such a way that, after removal of all edges,  a point array is obtained that has the same rotational symmetries as the original pentagon. Since every point in the array is related to every other via application of generators of the extended group, all points can be generated from a single point via the action of the extended group. They are hence collectively encoded by the group structure. 

The points in the array correspond to words in the generators of the affine extended group. Thus, if points are located in more than one of the translated and rotated copies of the original pentagon, then these points, called coinciding points, correspond to  non-trivial relations between group elements, and the extended group is hence not the free group. The point set obtained in Fig.~\ref{figpent} with the  translation of length $\tau=\frac{1}{2}(1+\sqrt{5})$ has cardinality 25, as opposed to 30, which would be the value in the generic case. Translations giving rise to such coinciding points are hence distinguished from a group theoretical point of view. 

Note that the point array contains a composition of a  pentagon and a decagon of different scaling, both centred on the origin. The affine group determines their relative sizes (or radial levels), and hence introduces radial information in addition to that encoded by the original group structure. Affine symmetry therefore allows one to constrain the overall geometry of a multishell structure from just part of the blueprint. 



Having illustrated our rationale in the two-dimensional setting, we now consider the icosahedral group, which is the largest rotational symmetry group in three dimensions. The icosahedral group $I$ consists of 60 rotations, and has 15 axes of 2-fold rotational symmetry, 10 axes of 3-fold symmetry, and 6 axes of 5-fold symmetry. In analogy to the two-dimensional example above, affine extensions have been constructed in \cite{Keef:2009} via the introduction of translation operators. In this construction, the icosahedron (whose 12 vertices lie on the axes of 5-fold symmetry), the dodecahedron (whose vertices are located on the 3-fold axes), and the icosidodecahedron (whose vertices are positioned on the 2-fold axes) have been used as geometric representations of icosahedral symmetry, and the directions and lengths of the translation operators in the affine extension have been derived with reference to these. The reason for this choice of polyhedral shapes, called start configurations in \cite{Keef:2009}, stems from the fact that they correspond to the projections of the standard bases of the three Bravais lattice types with icosahedral symmetry in 6 dimensions: the icosahedron obtained from a 6D simple cubic lattice,  the dodecahedron from a 6D body centred cubic lattice, and the icosidodecahedron from a 6D face centred cubic lattice. In particular, the use of these shapes in the construction of the affine groups ensures that the point arrays are subsets of the vertex sets of quasilattices with icosahedral symmetry; see \cite{Keef:2013} for a 2-dimensional example and \cite{Salthouse:2013} for a 3-dimensional one. In particular, this implies that the affine extended non-crystallographic groups are by construction related to aperiodic tilings, in analogy to the relation between affine extended crystallographic groups and lattices. 

\begin{figure}
\begin{center}
\includegraphics[height=5cm]{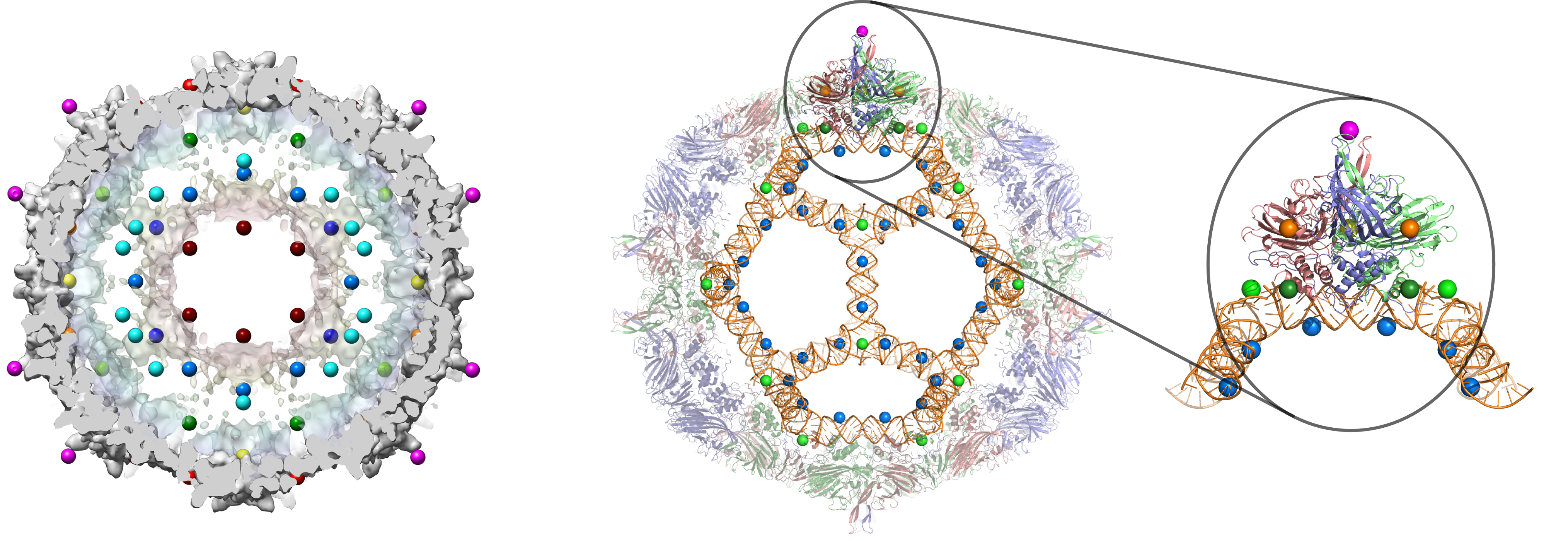}
\end{center}
\caption{A novel scaling principle in the structure of Pariacoto virus. The point arrays derived from affine extensions of the icosahedral group constrain different components of the overall viral geometry. In particular, they relate the RNA organisation of the packaged genome with structural features in the outer protein capsid, which were not previously thought to be related via symmetry.}
\label{Pariacoto}
\end{figure}

A full classification of the affine extensions of the icosahedral group based on the three polyhedral start configurations given by the icosahedron, dodecahedron and icosidodecahedron has been provided in \cite{Keef:2009}, and applications of these point arrays to viruses have been discussed in \cite{Keef:2013}. In particular, since the relative scalings between all the points in the array are fixed by the extended group, there is only one global scaling factor that maps all points collectively onto the  biological system. For example, in the case of Pariacoto virus discussed in \cite{Keef:2013} and shown in Fig. \ref{Pariacoto}, the length scales are determined by the structure of the genomic RNA such that array points map into the minor grooves of the molecule. The fact that the overall geometry is given by an orchestrated interplay between the interior RNA structure and the outer protein capsid structure is very surprising, and hints that the underlying  symmetry is actually affine. 

We revisit this classification here in the context of carbon chemistry. In this case, length scales are determined by the distances between carbon atoms, and only those point arrays are relevant that contain outer shells corresponding to three-coordinated cage structures. In \cite{Keef:2009,Keef:2013} only affine extensions of the chiral icosahedral group $I$ were considered, since viruses  do not normally possess inversion invariance. In \cite{DechantTwarockBoehm2011H3aff,DechantTwarockBoehm2011E8A4} we were working in a Coxeter group framework and therefore used the full icosahedral group $H_3=I_h$. Since here we are considering start configurations that are invariant under the full $I_h$ rather than just $I$, as well as extending along axes of icosahedral symmetry (which are also invariant under the full $I_h$), the resulting point arrays will also be invariant under the full icosahedral group $I_h$. 

\begin{figure}
\begin{center}
  (a)\includegraphics[height=4.5cm]{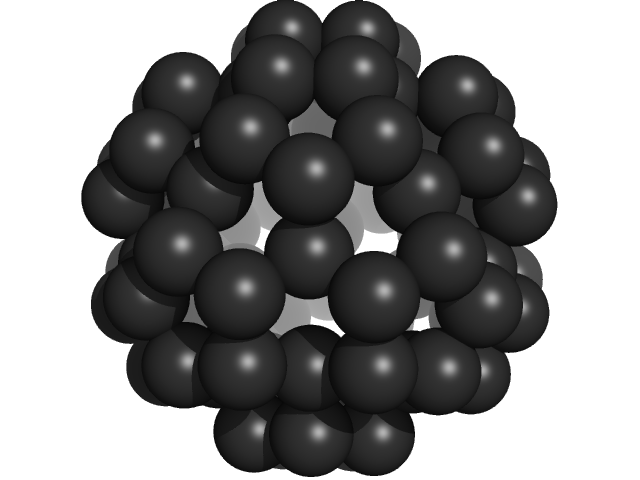}
  (b)\includegraphics[height=4.5cm]{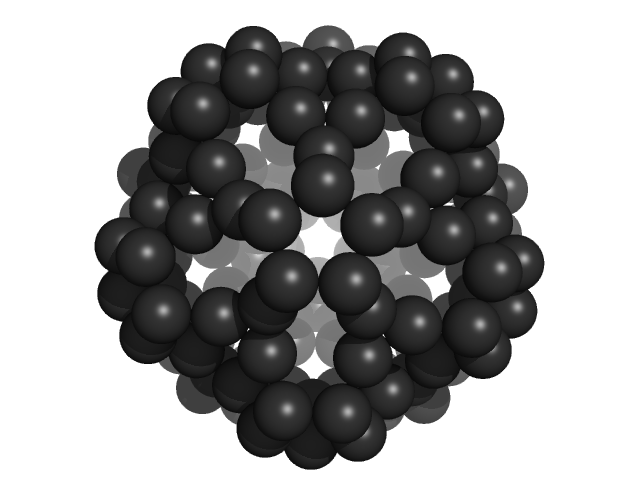}
  (c)\includegraphics[height=4.5cm]{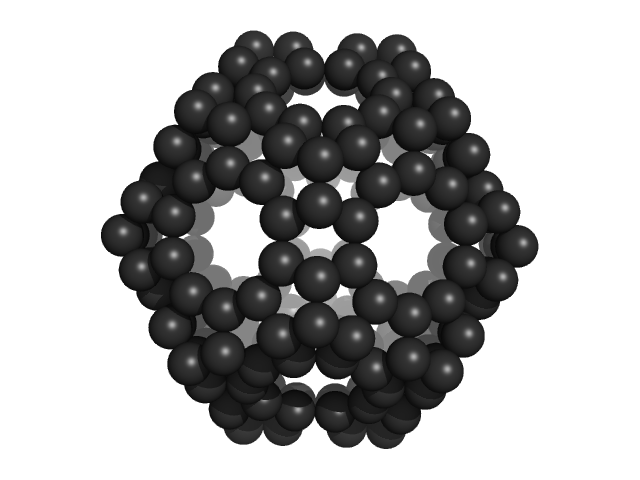}
\caption{The allowed translations for an icosahedral, dodecahedral and icosidodecahedral start configuration that yield trivalent configurations with 80, 120 and 200 vertices, respectively. Due to their non-uniform bonding structure, these configurations may not meet the constraints on bond angles and lengths required in carbon chemistry. }
\label{fig:Tom_3conn}
\end{center}
\end{figure}

For the dodecahedral start configuration, there are four affine extensions that result in three-coordinated shells. These are: a translation along  a 3-fold axis (lengths $\tau^2$ and $1/\tau^2$ yield the same overall structure) and along a 5-fold axis (length $1$ and $\tau$), with  200, 80 and 120 atomic positions, respectively, see Table~\ref{tab_allowedTom}. For the icosahedral start configuration, only a translation along a 3-fold axis of length 1 gives such a shell, which corresponds to the structure with 80 vertices already encountered in the dodecahedral case. For the icosidodecahedral start configuration, none of the affine extensions result in a three-coordinated shell.  The corresponding configurations are displayed in Fig.~\ref{fig:Tom_3conn}. Of these, only the shell with 80 atomic positions has approximately uniform angles between edges of the trivalent vertices, but  deviates from the structure published in \cite{Bodner2013dm5045} and the known structures of the $C_{80}$ fullerene isomers, suggesting that this structure  is  not realised in nature. 


\section{Application to fullerenes -- two families of carbon onions} \label{ACA_CarbonOnions}

Carbon onions are nested fullerene cages formed typically from three carbon cages of different size. Kustov \emph{et al.} \cite{Kustov:2008} have shown that from a group theoretical point of view, $C_n$ with $n=60z$ and $n=60z+20$, for $z\in\mathbb{N}$, are ``allowable'' icosahedral fullerene structures. We therefore extend our classification of affine extensions of the icosahedral group here to include start configurations corresponding to the cases $n=60$ and $n=80$ with a view to recover these carbon onions in our affine extension framework; note that the case of $n=20$ corresponds to the dodecahedron discussed in the previous section. 

\begin{figure}
\begin{center}
  (a)\includegraphics[height=4.5cm]{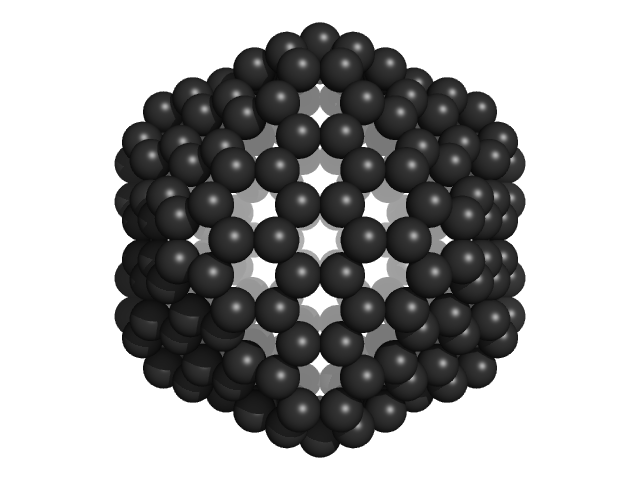}
  (b)\includegraphics[height=4.5cm]{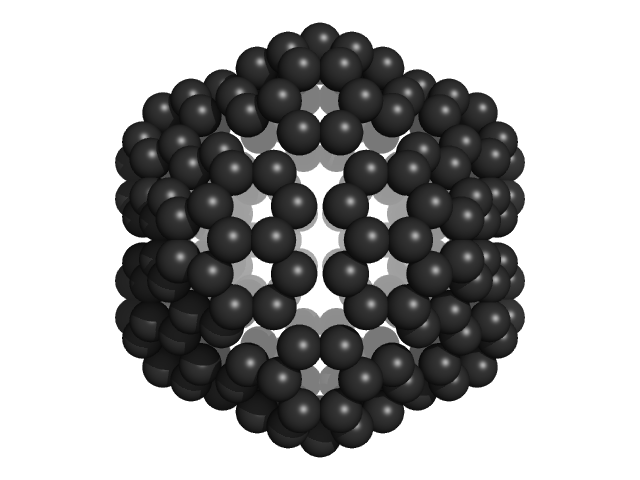}
  (c)\includegraphics[height=4.5cm]{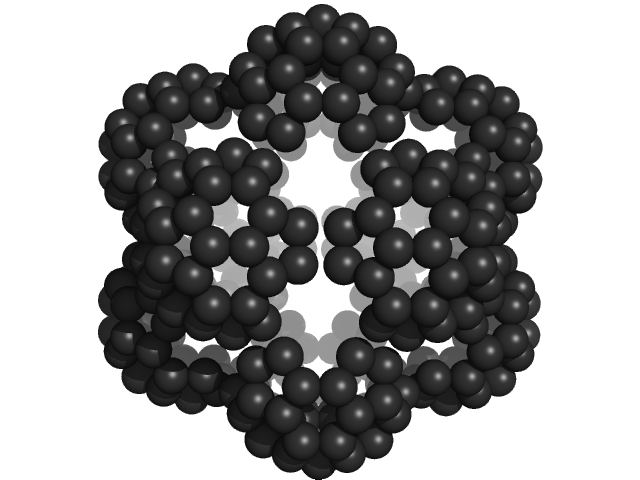}
\caption{The allowed translations for a start configuration corresponding to the structure of the $C_{60}$ fullerene cage, yielding trivalent configurations with 240, 240 and 360 vertices, respectively. Apart from the first configuration with 240 vertices, which is a good model for $C_{240}$ (panel (a)), these configurations are perhaps again not uniform enough in terms of bond angles and lengths to be physical models. }
\label{fig:C60_3conn}
\end{center}
\end{figure}

\begin{figure}
\begin{center}
  (a)\includegraphics[width=4cm]{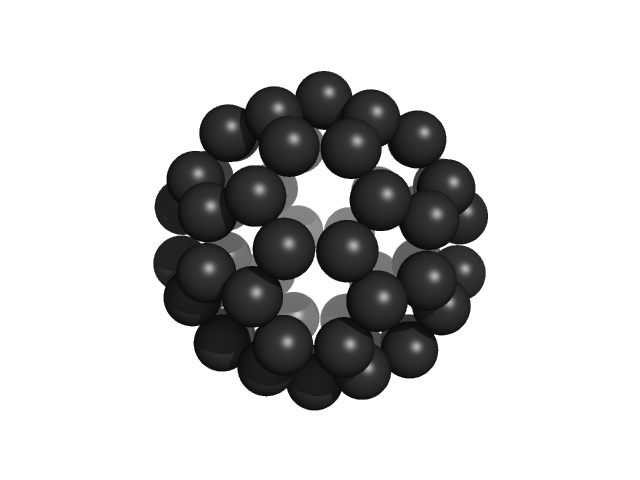}
  (b)\includegraphics[width=4cm]{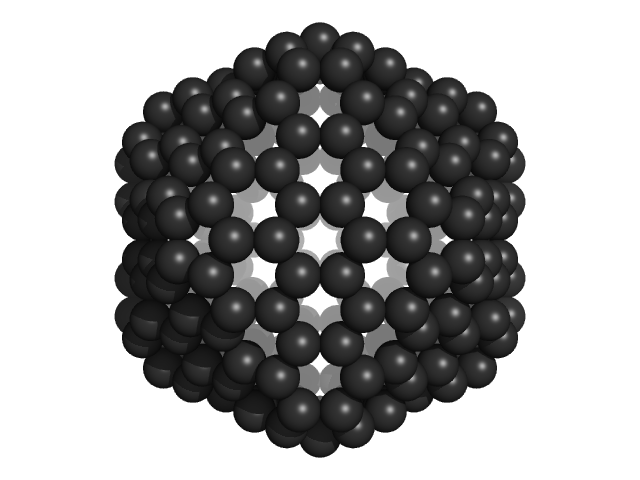}
  (c)\includegraphics[width=4cm]{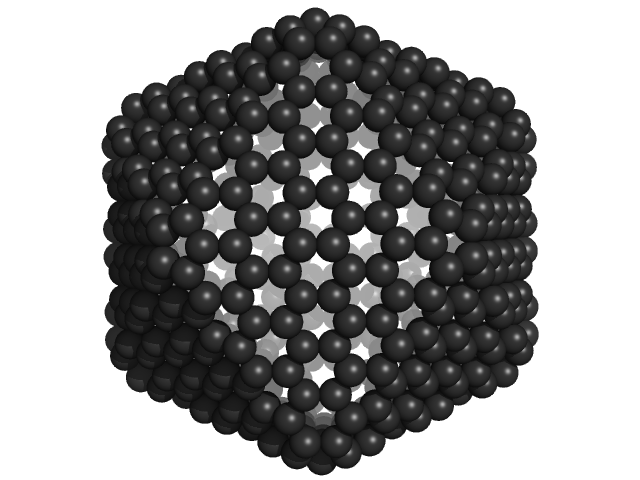}
\caption{The carbon onion generated from $C_{60}$: (a) $C_{60}$, (b) $C_{240}$ and (c) $C_{540}$. Note that all three have pentagons that are oriented vertex-to-vertex. Each iteration with the translation from the affine extension inserts an additional hexagon between the pentagons, thereby creating larger and larger shells.}
\label{fig:C60_C240_C540_pentagons}
\end{center}
\end{figure}

We start by applying the approach of affine extension to the truncated icosahedron, the cage structure corresponding to the fullerene $C_{60}$. The most natural coordinates for the truncated icosahedron include vertices of the form $(1,0, 3\tau)$, and vertices are hence located on a sphere with radius $\sqrt{10+9\tau}\approx 4.95$. We obtain a total of 49 nontrivial affine extensions. Among these, only three correspond to a three-coordinated cage. The details of these translations are given in Table \ref{tab_allowedC60} and the corresponding point array configurations are displayed in Fig.~\ref{fig:C60_3conn}. 
In particular, a translation along a 5-fold axis of length $3$ results in a point array whose outermost shell has $240$ three-connected vertices, positioned according to the structure of the fullerene $C_{240}$. Figure \ref{fig:C60_C240_C540_pentagons} shows how the structure of $C_{240}$ differs from that of $C_{60}$ by an extra hexagon 
 between the two pentagons, which are oriented vertex to vertex. 

Note that these point array configurations correspond to words in the generators of the extended groups in which the translation operations occurs precisely once. We therefore next generate the point arrays corresponding to words with precisely two occurrences of the translation operator. In particular, if a further iteration step is carried out for the translation that has generated the $C_{240}$ cage (i.e. another translation along a 5-fold axis with a multiplier of 3), a trivalent cage corresponding to the  structure of $C_{540}$ (shown in Figure \ref{fig:C60_C240_C540_pentagons}(c)) is obtained. It has one more extra hexagon between the pentagons as demonstrated in the figure. The fullerene cages $C_{60}$,  $C_{240}$ and $C_{540}$ are hence described by the same affine group, and the dimensions of all three shells are fixed relative to each other by this group. Interestingly, these three shells are known to occur collectively in nature in the form of a Russian doll like organisation called carbon onion; the nested carbon onion arrangements of such fullerenes are hence orchestrated by this affine symmetry group. Repeated application of this same affine extension generates a family of nested shells $C_{60}-C_{240}-C_{540}-\dots$, which in addition contain models for $C_{960}$, $C_{1500}$, $C_{2160}$ and $C_{2940}$ as well. 
This carbon onion is well-known experimentally \cite{Ugarte:1995}, and follows the pattern $C_n$ where $n$ is given by $n=60z^2$ for $z\in\mathbb{N}$. 
We have shown here that our affine symmetry approach describes different shells of such a carbon onion within a single framework.

\begin{figure}
\begin{center}
  (a)\includegraphics[width=4cm]{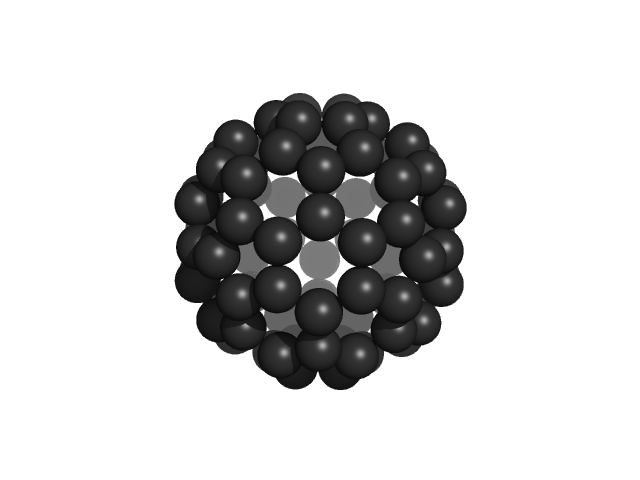}
  (b)\includegraphics[width=4cm]{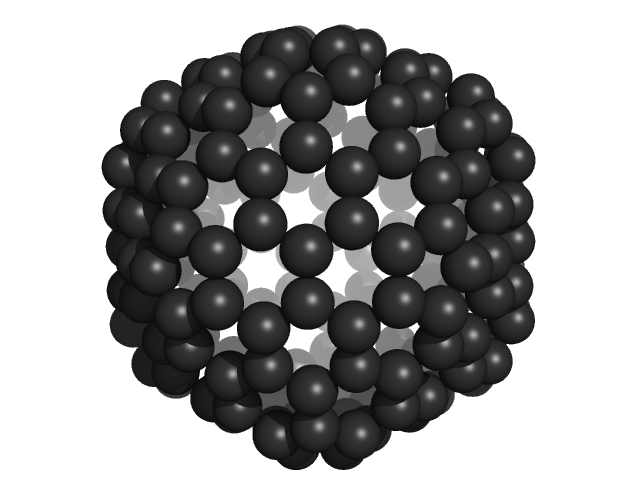}
  (c)\includegraphics[width=4.7cm]{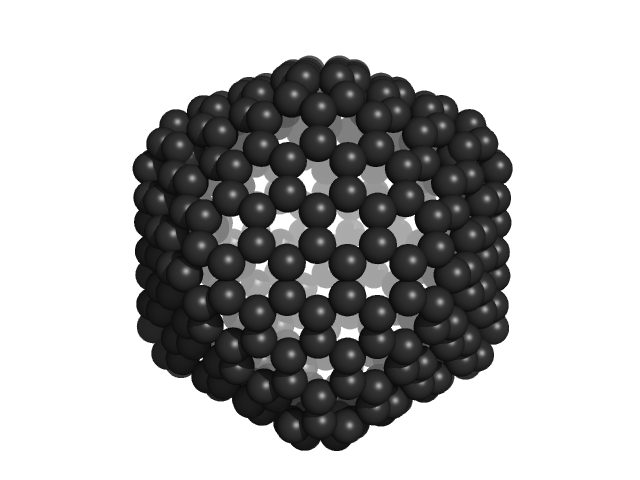}
\caption{The carbon onion generated from $C_{80}$: (a) $C_{80}$, (b) $C_{180}$ and (c) $C_{320}$. All three have pentagons that are oriented edge-to-edge, with each affine translation step inserting an extra hexagon between the pentagons.}
\label{fig:C20_C80_C180_pentagons}
\end{center}
\end{figure}

Next we start from the configuration $C_{80}$ in a parametrisation with radius $2\sqrt{3}\tau$, i.e. including vertices of the form $(2\tau, 2\tau, 2\tau)$. In contrast to the published isomers of lower symmetry, this configuration is chosen to have full $I_h$ symmetry. This start configuration yields 76 non-trivial affine extensions; however, only one of them (length $\frac{1}{5}(7+\tau)$ along an axis of 5-fold symmetry) corresponds to a three-connected outer shell. This shell consists of 180 vertices that are positioned according to the structure of a $C_{180}$ fullerene cage, see panel (b) of Fig. \ref{fig:C20_C80_C180_pentagons}. As before, affine extension has inserted an extra hexagon between two pentagons, however this time in an edge-to-edge conformation. We again consider words in the generators containing more than one copy of the translation operator, i.e. we repeat the copy-and-translate process using the same translation. This results in the successive insertion of further hexagons edge-on between the edge-on pentagons as shown in Figure \ref{fig:C20_C80_C180_pentagons}, thereby creating larger and larger shells corresponding to the fullerenes $C_{180}$, $C_{320}$, $C_{500}$, $C_{720}$ etc. These form the well-known carbon onion $C_{80}-C_{180}-C_{320}-\dots$ that follows a similar pattern $C_n$ where this time $n$ is given by $n=20(z+1)^2$ for $z\in\mathbb{N}$ \cite{terrones2002structure}.

We have thus found carbon onions based on the first two allowed icosahedral fullerene structures identified in \cite{Kustov:2008} within our affine extension framework. The next larger cage in their analysis has cardinality 120, and we therefore consider this case next. However, a similar analysis as above shows that there are no other three-connected cages arising via affine extension. We therefore terminate our analysis at this point. 


\section{Discussion and Conclusion}\label{ACA_Concl}

We have shown that affine icosahedral symmetry is ubiquitous in Nature. Apart from orchestrating the distribution of material at different radial levels in a virus, it models nested carbon cage structures such as fullerenes. 
We have derived here the two simplest carbon onions starting with $C_{60}$ and $C_{80}$ in a simple affine extension framework in which the affine symmetry relates all shells to each other simultaneously.
Carbon onions at higher orders such as ones starting at $C_{120}$ and $C_{140}$, which are the next allowed structures in the analysis of \cite{Kustov:2008}, are known, but are in fact chiral \cite{terrones2002structure}. \cite{terrones2002structure} further mention another chiral carbon onion starting with $C_{80}$, with $C_{240}$ as the next shell.
In this paper, we have considered extensions along the icosahedral symmetry axes. 
These extensions cannot yield chiral configurations; however, we have shown in \cite{DechantTwarockBoehm2011H3aff} that chiral point arrays can be obtained by extending along a direction other than an  axis of symmetry. 
We will therefore consider more general icosahedral configurations and translations in future work.

We note that a description of fullerene cages in terms of orbitals has been pioneered by Kustov et al (\cite{kustov2008theory,kustov2008molecular,kustov2009composition,kustov2012orbital}), and has provided a classification of fullerene architectures. 

We also note that an insertion of a hexagon between two pentagons in vertex-to-vertex orientation such that the resulting pentagons change to edge-on conformation (and vice versa) is  possible with the introduction of twist translations. Twist translations are analogues to glide reflections in higher dimensions, and are composites of translations along axes of $n$-fold rotational symmetry and rotations around these axes. Details of affinisations via twist translations are introduced and discussed further in \cite{Wardman:2012}. This work implies that the $C_{60}$ and $C_{80}$ configurations, that have been used here as start configurations, can indeed be generated from the icosahedral, dodecahedral and icosidodecahedral start configurations used in virus related work via twist translations. This again implies a link with the three Bravais lattice types in 6 dimensions, because -- as mentioned in the introduction -- these three start configuations can be obtained from the bases of these lattices via projection.

It is interesting to note that  we do not require either generalisation here for our purposes, as the simple affine extension framework is sufficient to derive the two most common  carbon onions in a simple, systematic, exhaustive and efficient manner. 
The idea of Caspar and Klug \cite{Caspar:1962} of explaining icosahedral virus structure by curving a planar hexagonal lattice into an icosahedron and inserting pentagonal defects at the corners has been extended to the fullerene case, see for instance \cite{terrones2002structure}. 
However, in the virus case we have shown that our approach in terms of Viral Tiling Theory and affine extensions of Coxeter groups is more general \cite{Twarock:2005a,Twarock:2004a} than the Caspar-Klug approach and hence provides additional insights into virus architecture.
The framework presented here is therefore also likely to be better suited to similar questions arising in the context of fullerenes.

\ack{Reidun Twarock would like to express her gratitude for a Leverhulme Research Leadership Award, which has provided funding for Dechant, Keef and Wardman.}




		%
		\begin{table}
		\begin{centering}\begin{tabular}{|c|c|c|c|c|c|}
		\hline
		\hline
	Polytope&	Direction&Length&Vertices
		\tabularnewline 	\hline 	\hline

	Dodec	&3&$\tau^{-2}$ or $\tau^2$&200
		\tabularnewline	\hline
		&	5&$\tau$&120
			\tabularnewline
		\hline
		&5&$1$&80
		\tabularnewline
			\hline
		Icos	&3 &$\tau-1$&120
			\tabularnewline
				\hline
				&3 &$1$&80
				\tabularnewline
			\hline
		\end{tabular}\par\end{centering}
		\caption
		{\label{tab_allowedTom} The translations that generate affine extensions of icosahedral symmetry resulting in  three-connected shells for start configurations given by an icosahedron of radius $\sqrt{\tau+2}$, a dodecahedron of radius $\sqrt{3}$ and an icosidodecahedron of radius  $\tau$.  }
		\end{table}
		%

	%
	\begin{table}
	\begin{centering}\begin{tabular}{|c|c|c|c|c|c|}
	\hline
	\hline
	Direction&Length&Vertices
	\tabularnewline 	\hline 	\hline
	5&$3$&240
	\tabularnewline	\hline
	5&$2\tau$&240
	\tabularnewline
		\hline
		5&$3\tau$&360
		\tabularnewline
	\hline
	\end{tabular}\par\end{centering}
	\caption
	{\label{tab_allowedC60} The translations that generate affine extensions of icosahedral symmetry for the start configuration corresponding to the cage structure of the $C_{60}$ fullerene and yield trivalent outer shells. }
	\end{table}
	%


\begin{figure}
\begin{center}
\includegraphics[height=5cm]{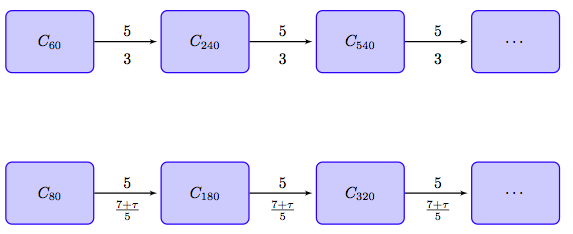}
\end{center}
\caption{The two carbon onions generated from $C_{60}$ and $C_{80}$, parametrised such that vertices are positioned at radial levels $\sqrt{10+9\tau}$ and $2\sqrt{3}\tau$, respectively. These are generated via translations along a 5-fold axis with length 3 in the former, and $\frac{1}{5}(\tau+7)$ in the latter case. }
\label{figonion}
\end{figure}

\end{document}